Journal Pre-proof

Present criteria for prophylactic ICD implantation: Insights from the EU-CERT-ICD (Comparative Effectiveness Research to Assess the Use of Primary ProphylacTic Implantable Cardioverter Defibrillators in EUrope) project

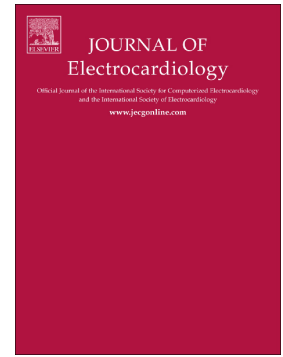

Markus Zabel, Simon Schlögl, Andrzej Lubinski, Jesper Hastrup Svendsen, Axel Bauer, Elena Arbelo, Sandro Brusich, David Conen, Iwona Cygankiewicz, Michael Dommasch, Panagiota Flevari, Jan Galuszka, Andreas Katsimardos, Jim Hansen, Gerd Hasenfuß, Robert Hatala, Heikki V. Huikuri, Tuomas Kenttä, Tomasz Kuczejko, Helge Haarmann, Markus Harden, Svetoslav Iovev, Stefan Kääb, Gabriela Kaliska, Andreas Katsimardos, Jaroslaw D. Kasprzak, Dariusz Qavoq, Lars Lüthje, Marek Malik, Tomáš Novotný, Nikola Pavlović, Peter Perge, Christian Röver, Georg Schmidt, Tchavdar Shalganov, Rajeeva Sritharan, Martin Svetlosak, Zoltan Sallo, Janko Szavits-Nossan, Vassil Traykov, Bert Vandenberk, Vasil Velchev, Marc A. Vos, Stefan N. Willich, Tim Friede, Rik Willems, Béla Merkely, Christian Sticherling, the EU-CERT-ICD Study Investigators

Please cite this article as: M. Zabel, S. Schlögl, A. Lubinski, et al., Present criteria for prophylactic ICD implantation: Insights from the EU-CERT-ICD (Comparative Effectiveness Research to Assess the Use of Primary ProphylacTic Implantable Cardioverter Defibrillators in EUrope) project, *Journal of Electrocardiology*(2018), https://doi.org/10.1016/j.jelectrocard.2019.09.001

This is a PDF file of an article that has undergone enhancements after acceptance, such as the addition of a cover page and metadata, and formatting for readability, but it is not yet the definitive version of record. This version will undergo additional copyediting, typesetting and review before it is published in its final form, but we are providing this version to give early visibility of the article. Please note that, during the production process, errors may be discovered which could affect the content, and all legal disclaimers that apply to the journal pertain.

**Present criteria for prophylactic ICD implantation:**

**Insights from the EU-CERT-ICD (Comparative Effectiveness Research to Assess the Use of Primary**

**ProphylacTic Implantable Cardioverter Defibrillators in EUrope) project**

Markus Zabel[1,2] MD, Simon Schlögl[1] MD, Andrzej Lubinski[3] MD, Jesper Hastrup Svendsen[4] MD DMSc,

Axel Bauer[5] MD, Elena Arbelo[6] MD, Sandro Brusich[7] MD, David Conen[8,9] MD, Iwona Cygankiewicz[10] MD,

Michael Dommasch[11] MD, Panagiota Flevari[12] MD, Jan Galuszka[13] MD, Andreas Katsimardos[12] MD,

Jim Hansen[14] MD, Gerd Hasenfuß[1,2] MD, Robert Hatala[15] MD, Heikki V. Huikuri[16] MD, Tuomas Kenttä[16] MD,

Tomasz Kuczejko[3] MD, Helge Haarmann[1] MD, Markus Harden[17], MSc, Svetoslav Iovev[18] MD, Stefan Kääb[5] MD,

Gabriela Kaliska[19] MD, Andreas Katsimardos[12] MD, Jaroslaw D. Kasprzak[20] MD, Dariusz Qavoq[20] MD,

Lars Lüthje[1] MD, Marek Malik[21] PhD MD, Tomáš Novotný[22] MD, Nikola Pavlović[23] MD, Peter Perge[24] MD,

Christian Röver[17] PhD, Georg Schmidt[11] MD, Tchavdar Shalganov[25] MD, PhD, Rajeeva Sritharan[1] MSc,

Martin Svetlosak[15] MD, Zoltan Sallo[24] MD, Janko Szavits-Nossan[26] MD, Vassil Traykov[27] MD,

Bert Vandenberk[28] MD, Vasil Velchev[29] MD, Marc A. Vos[30] PhD, Stefan N Willich[31] MD,

Tim Friede[2,17] PhD, Rik Willems[28] MD PhD, Béla Merkely[24] MD, PhD, Christian Sticherling[8] MD,

for the EU-CERT-ICD Study Investigators

[1]Dept. of Cardiology and Pneumology, Heart Center, University Medical Center, Göttingen, Germany; [2]DZHK

(German Center for Cardiovascular Research), partner site Göttingen, Göttingen, Germany; [3]Dept. of

Cardiology, Medical University of Lodz (MUL) WAM Hospital, Lodz, Poland; [4]Dept. of Cardiology, The Heart

Centre, Rigshospitalet, Copenhagen University Hospital, and Department of Clinical Medicine, University of

Copenhagen, Copenhagen, Denmark, [5]Dept. of Cardiology, Klinikum Großhadern, Ludwig-Maximilians-

Universität Munich, Germany; [6]IDIBAPS, Dept. of Cardiology, Hospital Clinic Barcelona, Spain; [7]Dept. of cardiovascular disease, KBC Rijeka, Rijeka, Croatia, [8]University Hospital, University of Basel, Switzerland; [9]Population Health Research Institute, McMaster University, Hamilton, Ontario, Canada; [10]Dept. of Cardiology, Medical University of Lodz (MUL) CKD Hospital, Lodz, Poland; [11]Med. Klinik und Poliklinik I, Technische Universität München, Klinikum rechts der Isar, Munich, Germany; [12]2nd Dept. of Cardiology, Attikon University Hospital, Athens, Greece; [13]Dept. of Cardiology, University Hospital and Faculty of Medicine and Dentistry, Palacký University, Olomouc, Czech Republic; [14]Gentofte Hospital, Copenhagen, Denmark; [15]Slovak Medical University and NUSCH, Bratislava, Slovakia; [16]Medical Research Center, Oulu University Hospital and University of Oulu, Finland; [17]Dept. of Medical Statistics, University Medical Center Göttingen, Göttingen, Germany; [18]Dept. of Cardiology, St. Ekaterina University Hospital, Sofia, Bulgaria; [19]Dept. of Cardiology, SUSSCH Banska Bystrica, Slovakia; [20]Chair and Dept. of Cardiology, Bieganski Hospital, Medical University of Lodz (MUL), Lodz, Poland; [21]National Heart and Lung Institute, Imperial College, London, United Kingdom; [22]Dept. of Internal Medicine and Cardiology, University Hospital Brno, Masaryk University, Brno, Czech Republic; [23]Dept. of Cardiology, KBC Sestre Milosrdnice, Zagreb, Croatia; Heart and Vascular Center, Semmelweis University , Budapest/Hungary [25]Dept. of Cardiology, National Heart Hospital, Sofia, Bulgaria; [26]Dept. of Cardiology, Magdalena Klinika, Krapinske Toplice, Croatia; [27]Dept. of Cardiology, Acibadem City Clinic Tokuda Hospital; [28]University Hospitals of Leuven, Leuven, Belgium; [29]Dept. of Cardiology, St. Anna Hospital, Sofia, Bulgaria; [30]Dept. of Medical Physiology, University Medical Center Utrecht; [31]Institute for Social Medicine, Epidemiology and Health Economics, Charité - Universitätsmedizin Berlin, Berlin, Germany;

**Running title:** Present criteria for primary prophylactic ICD implantation

**Abstract word count:** 242; **Word count:** 2463

***Address for correspondence:***

Markus Zabel, MD
Dept. of Cardiology and Pneumology,
Div. of Clinical Electrophysiology, Heart Center
University Medical Center
Robert-Koch-Strasse 40, 37075 Göttingen/Germany
Phone: +49 (551) 39 69255; Fax: +49 (551) 39 19127
E-Mail: markus.zabel@med.uni-goettingen.de

**Abbreviations**

AV = atrioventricular

CI = 95% confidence interval

CRT-D = cardiac resynchronization therapy defibrillator

CRT-P = cardiac resynchronization therapy pacemaker

DCM = dilated cardiomyopathy

ECG = electrocardiogram

EP = electrophysiologic

ESC = European Society of Cardiology

EU = European Union

EU-CERT-ICD = EUropean Comparative Effectiveness Research to Assess the Use of Primary ProphylacTic Implantable Cardioverter Defibrillators

GCP = Good Clinical Practice

HR = hazard ratio

ICD = implantable cardioverter defibrillator

ICM = ischemic cardiomyopathy

LVEF = left ventricular ejection fraction

NYHA = New York Heart Association

QoL = quality of life

SCD = sudden cardiac death

SPRM = Seattle Proportional Risk Model

US = United States

VA = Veterans Administration

**Abstract**

**Background.** The clinical effectiveness of primary prevention implantable cardioverter defibrillator (ICD) therapy is under debate. It is urgently needed to better identify patients who benefit from prophylactic ICD therapy. The EUropean Comparative Effectiveness Research to Assess the Use of Primary ProphylacTic Implantable Cardioverter Defibrillators (EU-CERT-ICD) completed in 2019 will assess this issue.

**Summary.** The EU-CERT-ICD is a prospective investigator-initiated non-randomized, controlled, multicenter observational cohort study done in 44 centers across 15 European countries. A total of 2327 patients with heart failure due to ischemic heart disease or dilated cardiomyopathy indicated for primary prophylactic ICD implantation were recruited between 2014 and 2018 (>1500 patients at first ICD implantation, >750 patients non-randomized non-ICD control group). The primary endpoint was all-cause mortality, first appropriate shock was co-primary endpoint. At baseline, all patients underwent 12-lead ECG and Holter-ECG analysis using multiple advanced methods for risk stratification as well as documentation of clinical characteristics and laboratory values. The EU-CERT-ICD data will provide much needed information on the survival benefit of preventive ICD therapy and expand on previous prospective risk stratification studies which showed very good applicability of clinical parameters and advanced risk stratifiers in order to define patient subgroups with above or below average ICD benefit.

**Conclusion.** The EU-CERT-ICD study will provide new and current data about effectiveness of primary prophylactic ICD implantation. The study also aims for improved risk stratification and patient selection using clinical risk markers in general, and advanced ECG risk markers in particular.

Keywords: implantable cardioverter defibrillator; risk factors; mortality; sudden cardiac death

## Introduction

Multicenter landmark trials have long established that implantable cardioverter-defibrillator (ICD) therapy is useful for primary prevention of sudden cardiac death (SCD)(1, 2). After guideline implementation, ICDs quickly became routine treatment. Almost two decades later, outcome events have noticeably decreased(3) and non-sudden modes of death compete with the occurrence of malignant arrhythmias and SCD so that many ICD patients never receive appropriate shocks(4). Important sub-groups, such as patients with more advanced heart failure(2, 5), with advanced kidney disease(6), women(7), or diabetics(8) may have an ICD benefit below average. The overall negative results of the DANISH-ICD trial(9) highlighted this dilemma and prompted further debate. DANISH showed that ICD therapy currently does not reduce mortality in all patients with non-ischemic cardiomyopathy. From the original DANISH data, it was shown that increasing age is associated with loss of ICD survival benefit(10), i.e. that age should be an important determinant of ICD indication. Better diagnostic criteria for indicating prophylactic ICD therapy are urgently needed to ensure the expected survival benefit from device therapy to the patient.

## Considerations on ICD survival benefit

Clinical risk scores to predict all-cause mortality in ICD candidates are ready to use with high accuracy and reproducibility(11). The incidence of malignant arrhythmias and appropriate ICD shocks is not proportional to overall mortality(5, 12) (see Figures 1, 2), therefore risk of arrhythmic death can vary greatly among patients with reduced left ventricular ejection fraction (LVEF). ICD survival benefit can be estimated by the relative magnitudes of malignant arrhythmia risk versus overall mortality risk(12-14), first demonstrated by Goldenberg et al(13) and Levy et al(14) (see Figures 1,3, adapted from (5)). Unfortunately, specific prediction of malignant ventricular arrhythmias, i.e. also SCD, is more difficult and requires different predictor variables(5, 12), for instance electrophysiologic (EP) stimulation. ICDs reliably abort SCD caused by ventricular tachycardia and ventricular fibrillation, so prediction of arrhythmic risk is critical to identify patients with positive ICD benefit. Using simultaneous prediction of expected mortality risk and appropriate shock, individual ICD benefits can be estimated (see Figures 1,3, adapted from(5)). Patients with very high mortality may not derive ICD benefit because of a higher proportion of non-sudden and non-cardiac death risk(4, 5, 12-14)(see Figures 1-3). If higher shock risk coincides with lower mortality risk, ICD survival benefit will be particularly high. As shown in Figure 3B (yellow curve), 20% of the patients in the EUTrigTreat clinical study (the upper quintile of the shock score) featured a ≈10% annual risk of appropriate shock, therefore they undoubtedly derive a high survival benefit from their device(5). There is a lower end of risk where event probabilities are too low for the defibrillator to exert benefit(13). In other circumstances of SCD risk, a <1% annual risk of SCD has been discussed as not sufficiently high for ICD prophylaxis. From the EUTrigTreat prospective cohort(5), a considerable number of ICD patients can be identified with low shock probability and low mortality, i.e. a low chance of benefit (Figures 1,3). We also found that 20% of patients in the EUTrigTreat study(5) featured a high all-cause mortality of ≈11% annually (Figure 3A: yellow curve). One cannot be exactly sure whether the high mortality in this subgroup confers a reasonable proportion of arrhythmic deaths which would make ICD therapy very useful or whether there is rather a high proportion of non-arrhythmic deaths among these patients limiting its use. This is where separate prediction of arrhythmic deaths and non-sudden deaths is urgently needed. EP testing is an excellent diagnostic test to predict arrhythmic risk(5), for practical reasons an ECG parameter with similar predictive capability would be highly desirable. Altogether, the scoring system derived from the data of Bergau et al (5) shows a good discrimination between low, moderate and high risks for appropriate shocks and all-cause mortality as indicated by C-statistics between 0.69 (shocks) and 0.86 (mortality). A similar grading system, a bimodal survival and implantable defibrillator (BaSIS) risk model was developed and proposed by the Ontario ICD registry group and Lee et al(15). In this paper, a meticulous collection of clinical information in 3445 patients undergoing ICD implantation accurately allowed to grade the two outcomes shock and death in these patients. The cohort was divided into risk deciles with very large differences in the risks of mortality (36-fold from first to tenth decile) and appropriate shock (8-fold from first to tenth decile). No additional diagnostic testing for risk stratification was undertaken as the basis for the study was a registry collection of data. In this respect, appropriate shock risk can be replaced by the proportional risk of sudden cardiac death which is the principle of the Seattle Proportional Risk Model (SPRM)(12). As a recent example, Kristensen et al measured the SPRM from the original DANISH study(16). In 1116 patients, the authors could show that 558 patients with an SPRM above the median had much better risk reduction by ICD therapy (HR 0.63, 95% CI 0.43 – 0.94) than 558 patients with SPRM below the median (HR 1.08, 95% CI 0.78 – 1.49, p for interaction = 0.04).

**Importance of age-specific ICD treatment effect**

The recent DANISH-ICD trial also showed that elderly patients with non-ischemic cardiomyopathy and an LVEF ≤ 35% did not derive a survival benefit from ICD treatment(9). In a post-hoc analysis from the DANISH study, age was demonstrated as an important factor connected to ICD benefit. ICD survival benefit was significant in patients younger than 70 years but was lost in elderly patients. The interaction p-value was 0.009 for age divided into tertiles which is a remarkable value for clinical studies of this size and duration(10). Harm could be suspected in patients above 80 years. These results are in contrast with meta-analytic data from older ICD landmark studies where a decline of ICD benefit with age was also shown(17), but patients up to an age of 90 years had a persistent survival benefit from ICD treatment. Cardiovascular treatments and outcomes may have widely changed between the two eras. From the Ontario ICD registry, Yung et al(18) had shown in 5399 patients that mortality increases with age but rates of appropriate shocks remain similar between all age groups. Sponsored by the US Veterans Administration the I-70 study (NCT 02121158) currently randomizes 1000 elderly patients ≥ 70 years to a primary prophylactic ICD vs. control without ICD.

**Rationale of European ICD studies**

The findings by Goldenberg et al(13) and Levy et al(14) were hypothesis-generating for the prospective EUTrigTreat(5, 19)(2009-2015) and EU-CERT-ICD studies(20)(2013-2018) which searched for the best combination of EP risk stratification tests and clinical parameters to predict competing risks of ICD shocks versus mortality and guide the selection for prophylactic ICD treatment. We hypothesized subgroups with high competing non-sudden or non-cardiac mortality as well as defined patients with a higher risk of malignant arrhythmias. In the EUTrigTreat clinical study, Bergau et al(5) showed that all-cause mortality risk and appropriate shock risk could be very accurately predicted in 672 ICD patients with primary prophylactic <u>and</u> secondary prophylactic indications (Figures 1-3). When designing the EU-CERT-ICD "<u>EU</u>ropean <u>C</u>omparative <u>E</u>ffectiveness <u>R</u>esearch to assess the use of primary prophylac<u>T</u>ic <u>I</u>mplantable <u>C</u>ardioverter <u>D</u>efibrillators (EU-CERT-ICD)" prospective study in 2012, a randomized trial did not seem ethically appropriate due to the wide implementation of ICD therapy and unequivocal guidelines. We therefore set out to design a large prospective study establishing a non-randomized control group without ICDs as feasible.

**The EU-CERT-ICD project – general concept**

In general, we made use of disparities of ICD implant rates across Europe(21). Several of the participating centers and countries did not have access to primary prophylactic implantation due to lack of reimbursement in some or all of their patients, these centers eventually recruited about 60% of the non-ICD control patients.

Furthermore, there were control patients that had refused a recommended primary prevention ICD implantation on personal preference. The study was funded by the European Community's Seventh Framework Programme (FP7). A central prospective study(20), a retrospective registry(7) and meta-analyses in primary prophylactic ICD patients were set up(22, 23). Over the course of the project so far, several original papers and editorials have been published from the various work packages (see www.eu-cert-icd.eu). Primary objectives of the project were 1) to characterize all-cause mortality in a prospective controlled non-randomized study of ICD candidates for primary prophylaxis of SCD, and compare newly implanted ICD patients with a non-randomized control group without ICD treatment; 2) to determine the contribution of prespecified clinical baseline characteristics to the risk of the primary outcomes; 3) to define subgroups within the ICD primary prevention guideline cohort with a lower or higher benefit from ICD treatment; 4) to assess simple and cost-effective electrocardiographic noninvasive risk stratification techniques; 5) to identify predictors for appropriate shocks using ECG-related parameters and autonomic parameters, and to characterize subgroups with a deviating risk for appropriate shock, in particular focusing on the role of sex category; 7) to biobank a genetic blood sample from each patient; 8) to provide outcome data for health economic evaluation of ICD use including quality of life (QoL) with particular focus to subgroups and country-specific differences;

**EU-CERT-ICD: Prospective study design and protocol**

The EU-CERT-ICD prospective trial (www.clinicaltrials.gov, NCT02064192) is an investigator-initiated non-randomized, open, controlled, observational multicenter study(20) in 2250 analyzable patients with ischemic or dilated cardiomyopathy that were candidates for a primary prevention ICD by current guidelines. In the ICD treatment group, 1500 analyzable patients at first ICD implantation were targeted. A non-randomized control group of 750 patients without ICDs was recruited to generate contemporary comparative data on ICD survival benefit. To achieve this, 2327 patients were targeted. Inclusion and exclusion criteria are shown in Table 1. Sample size calculations were done in order to compare ICD patients with controls regarding mortality, and for stratification of the ICD cohort for appropriate shocks and mortality. Estimated from previous studies, an annualized mortality of 4.5% and a first appropriate ICD shock rate of 4.5% were expected. For a power of 80% at a two-sided significance level of 5%, in a 4-year study a sample of 1500 ICD patients and 750 controls was calculated to generate at least 279 death events(20). For independent binary or dichotomized risk stratifiers to provide hazard ratios of about 2 between a high and a low risk group we calculated that 122 deaths yielded 95% power for a two-sided significance level of 5%. Correspondingly, 108 appropriate ICD shocks were required if the ratio of high and low risk group sizes was 1:1 and an annual appropriate ICD shock rate of about 4.5% was observed. Predefined subgroups were the elderly, diabetics, women and patients with comorbidities with a higher likelihood of non-arrhythmic death as also identified from the mortality score. Reasons for non-ICD status had to be unrelated to the study. Balancing of clinical characteristics between ICD and control groups was approached using multivariate analyses and propensity score methods. Primary endpoint was all-cause mortality, first appropriate shock was a co-primary endpoint in ICD patients. Secondary endpoints of the study included SCD, non-cardiac mortality, first inappropriate shock. A 12-lead Holter ECG (Getemed, Teltow/Germany) was recorded at 1kHz for 24 hours at baseline and analyzed for all standard ECG parameters, short-term advanced parameters, and advanced Holter ECG parameters (Table 2). The number of patients with atrial fibrillation was limited to 15%. A large number of clinical baseline parameters were documented at enrollment of each patient(20). ICD patients were followed every 3 to 6 months or remotely. Patients in the non-ICD control group were scheduled every 6 to 12 months according to clinical needs. Mandatory programming called for rate cutoffs at 200 and 240 bpm, programming changes were recorded. Outcome information was retrieved by phone and mail from patients, relatives, general practitioners, hospital records, or local authorities. Deaths were classified as SCD, cardiac, or non-cardiac. ICD shocks were adjudicated as appropriate or inappropriate. Crossover of patients from the control group to the ICD group was allowed at the discretion of the treating physicians, and occurred in ≈ 4% of patients who remained in the study. QoL was assessed at baseline and annually (SF-36, MacNew, Florida Patient Acceptance Survey). Validated health economics questionnaires were collected in Germany and Switzerland to assess true health care costs. QoL-adjusted cost-effectiveness is estimated from cost comparisons and Markov models. A combination of central monitoring and on-site monitoring was chosen to ensure data completeness, data quality and study conduct in accordance with the protocol and GCP guidelines. Cox proportional hazards regression analyses or Fine & Gray proportional sub-distribution hazard regression analyses are performed to quantify the predictive value of multiple categorical variables and dichotomized continuous variables.

**Other ICD studies with potential to change ICD indications**

The DO-IT registry study funded by Dutch health insurers will report outcomes including cost-effectiveness from 1500 primary prophylactic ICD patients with LVEF ≤ 35% from the Netherlands (24). The randomized

RESET-CRT trial (NCT03494933) is funded by the German health system and randomizes 2000 heart failure patients with an LVEF ≤ 35% and QRS ≥ 150 ms between CRT-D and CRT-P. The I-70 study sponsored by the VA (NCT 02121158) randomizes 1000 elderly patients >= 70 years to a primary prophylactic ICD vs. control without ICD. A number of studies are underway to evaluate risk stratification of ICM patients with LVEF between 36% and 50%. Overall average risk is smaller in this group, not excluding that selected patients have a high risk of SCD and may be good candidates for ICD treatment. These are REFINE-ICD (NCT00673842) using T-wave alternans and heart-rate turbulence, SMART-MI (NCT02594488) using periodic repolarization dynamics(25), PRESERVE EF (NCT02124018) using programmed ventricular stimulation, and CMR-Guide (NCT01918215) using magnetic resonance imaging.

**Summary and conclusions**

Appropriate identification of patient subgroups with significant mortality benefit from ICD therapy remains critical, and incorporation of variables beyond LVEF and NYHA functional class is warranted. It is almost certain that not all patients identified by LVEF ≤ 35% have survival benefit from ICD therapy. The EU-CERT-ICD study results upcoming in 2019 will provide contemporary data on effectiveness of primary prophylactic ICD implantation in ICM and DCM. Using multivariable regression statistics of the primary endpoint, we will calculate an adequately powered hazard ratio of the ICD survival effect (as the primary measure of ICD benefit) in the overall cohort and predefined subgroups. Risk scores for mortality and shock will be provided building on a large number of useful parameters, e.g. ECG markers, cardiovascular history, biomarkers, and possible combinations thereof. The results will permit assessment of the predictive value of several state-of-the-art advanced ECG methods for application in clinical decision-making. Facing a risk continuum of arrhythmic risk and SCD risk which has a wide range of proportions to all-cause mortality and cardiac mortality, patient decisions should be individualized and currently based on simple parameters such as age, NYHA, BNP, atrial fibrillation, creatinine, the existence of comorbidities, possibly adjunct diagnostic tests such as EP stimulation and magnetic resonance imaging may be used.

**FUNDING**

The research leading to the results has received funding from the European Community's Seventh Framework Programme FP7/2007–2013 under grant agreement No. HEALTH-F2-2009-241526, EUTrigTreat, and grant agreement No. HEALTH-F2-2009-602299, EU-CERT-ICD. G.H. and T.F. are principle investigators of the German

Center for Cardiovascular Research (DZHK), partner site Göttingen. R.W. is supported as a postdoctoral clinical

researcher by the Fund for Scientific Research Flanders (FWO).

**Conflicts of Interest**

none declared

**Figure legends**

**Figure 1: Distribution of patients to combinations of risk categories (low, intermediate, high) and their associated annualized mortality and shock risks (modified from reference (5) (EUTrigTreat clinical study), with permission)** Grey circles denote the frequencies of patients in the various categories. The red and blue bars denote the actual annualized shock and mortality risks in a category, respectively. For each risk, the cohort is divided into three risk groups (low: two quintiles, intermediate: two quintiles, high: one quintile), resulting in nine subgroups, of which seven have significant size. Annualized shock risk is found to be >10% per year in the highest quintile of the shock score and coincides with both an intermediate (4.4% per year) and a high (10.2% per year) mortality. Annualized mortality risk is found to be >10% per year in the highest quintile of the mortality score and coincides with both an intermediate (5.1% per year) and a high risk of appropriate shock.

**Figure 2: Correlation scatter plot for calculated risk score values of appropriate shock vs. calculated risk score value for mortality ($r = 0.56$, $p < 0.001$) (modified from reference (5) (EUTrigTreat clinical study), with permission)**

Horizontal and vertical lines depict the low, intermediate, and high-risk values of each score. The figure shows that the correlation is at best moderate despite statistical significance. Thus, all-cause mortality risk does not coincide well with appropriate shock risk. Horizontal and vertical lines show the classification of risk scores into low/intermediate/high categories. Individually, a low risk of appropriate shock does occur with a high competing risk of death limiting the effectiveness of implantable cardioverter defibrillator therapy in a given patient (lower right quadrant). Vice versa, individual patients can be identified with fairly high risks of appropriate shock and concomitant moderate risks of death (upper left quadrant). These individuals are expected to have a higher life-prolonging effect of their implantable cardioverter defibrillator therapy, i.e. higher implantable cardioverter defibrillator benefit.

**Figure 3: Differentiation of arrhythmia risk and all-cause mortality risk in ICD patients (modified from reference (5) (EUTrigTreat clinical study), with permission)**

Cumulative event-probability curves for mortality and appropriate shock (Panel A and B). For each risk, the cohort is divided into three risk groups (low: two quintiles, intermediate: two quintiles, high: one quintile), the calculation is provided by separate risk scores for all-cause mortality and appropriate shock. The dashed lines indicate the cumulative event-probabilities after bootstrap bias correction.

Panel A: The mortality risk score provides excellent separation of low, intermediate, and high mortality risks. The low risk mortality group (two quintiles) shows an annualized risk of 0.5%. In contrast, the high-risk mortality group (one quintile) features an annual risk of 11%. Within the latter patients, it can be expected that non-sudden cardiac deaths or noncardiac deaths compete with the occurrence of ventricular arrhythmias. Patients with a low predicted shock risk may not improve their prognosis wearing an ICD.

Panel B: The appropriate shock risk score provides good separation of low, intermediate, and high shock risks. The low risk shock group, a large group covering two quintiles (40% of patients in the cohort) has an average annual risk of 1.8%. Since a first appropriate shock corresponds with a potential SCD in 30–50%, this number corresponds to an SCD rate < 1%/year. In patients with an estimated SCD rate < 1% annually, depending on age and other mortality factors independent of arrhythmias, omission of an ICD may be discussed. In contrast, the high-risk group for shock (one quintile) features an average annual risk of ≈8.5%, qualifying the patient for an ICD with high survival benefit. In the intermediate risk of shock group (two quintiles), the risk is still ≈4% annually, corresponding to possibly a 2% annual SCD rate in a parallel non-ICD group. Patients in the intermediate risk group for shock should probably also obtain an ICD as they derive ICD benefit unless a very high competing risk

of non-arrhythmic mortality can be seen from the mortality score

**Figure 1:**

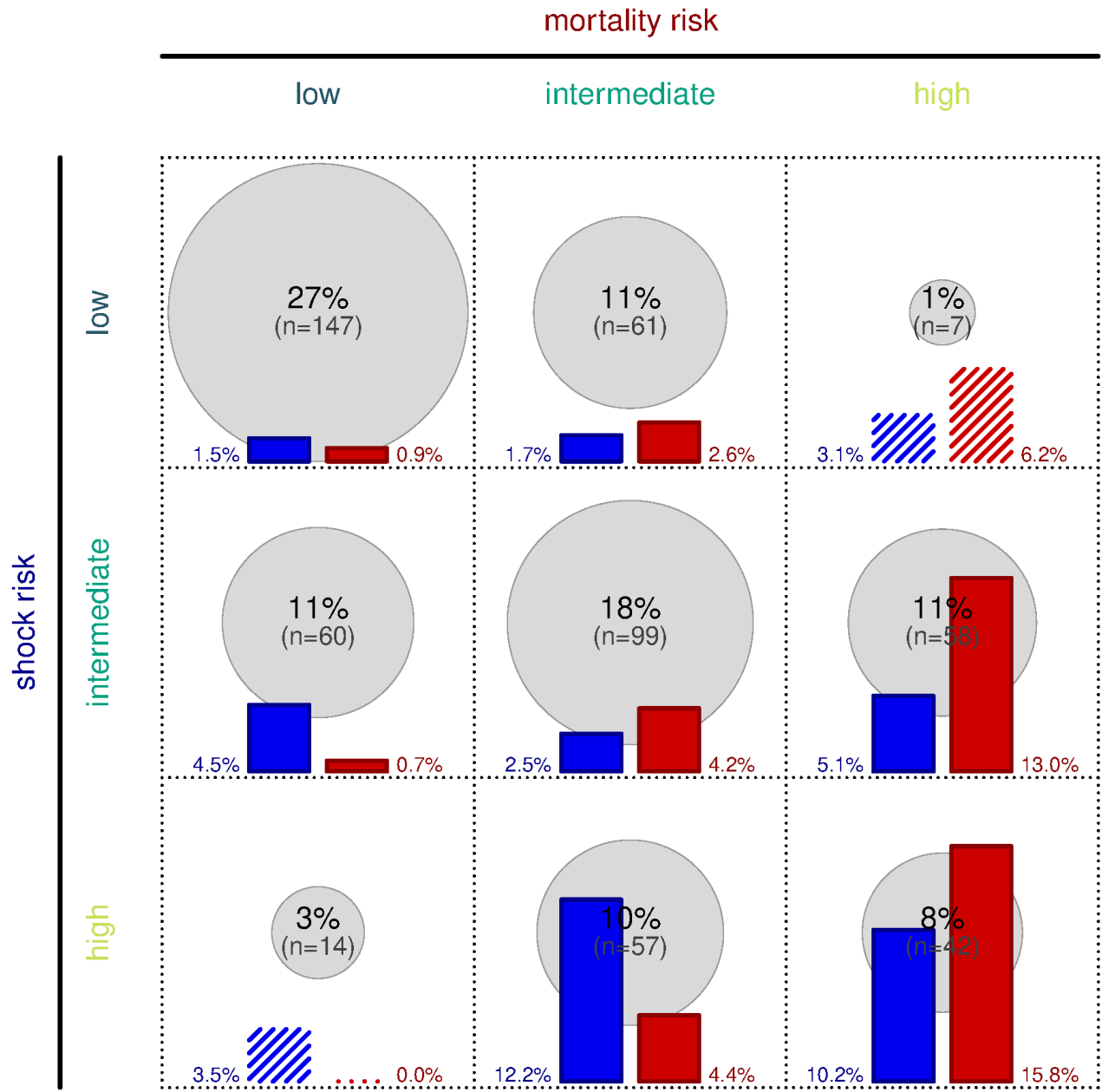

**Figure 2:**

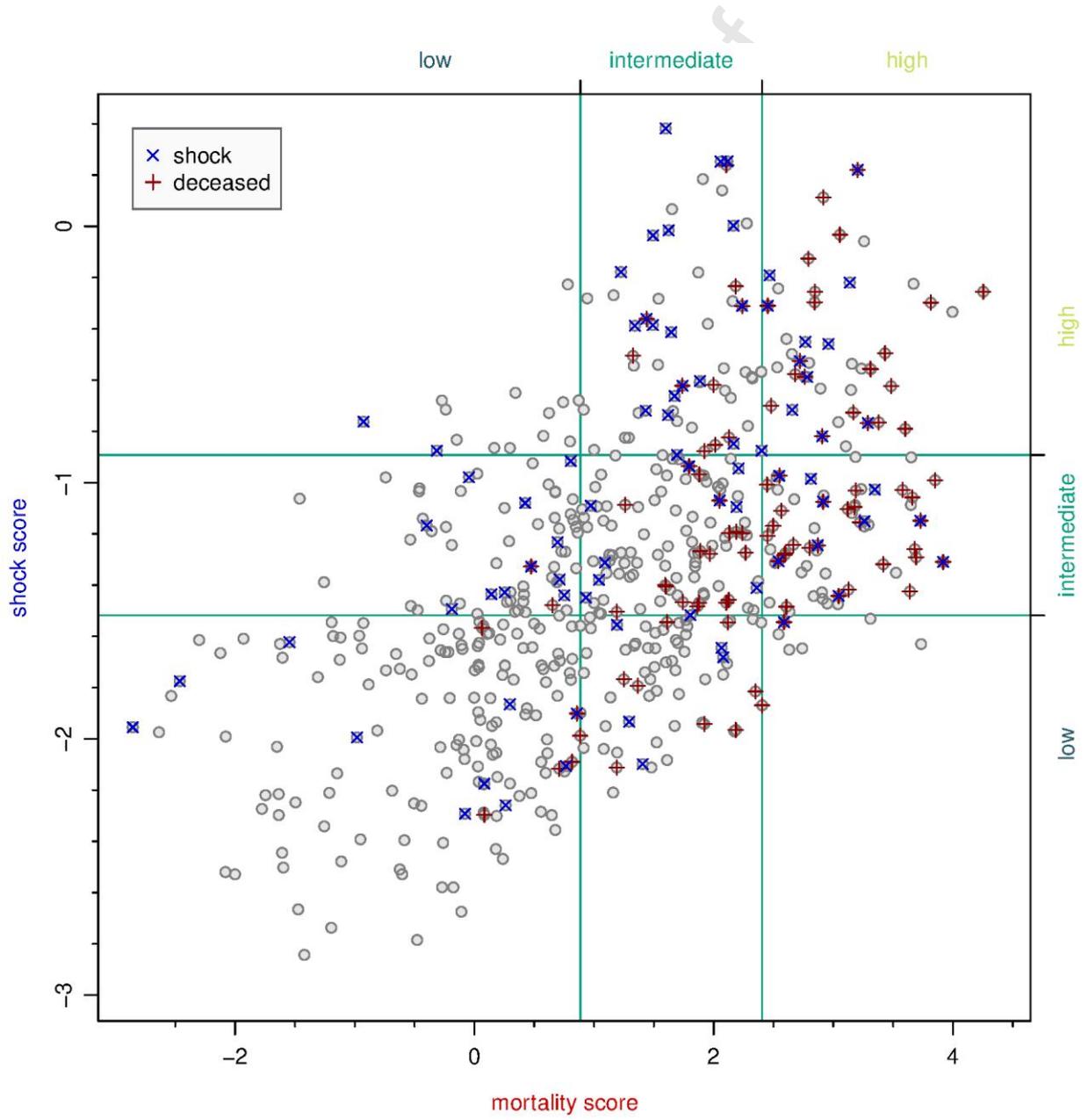

**Figure 3:**

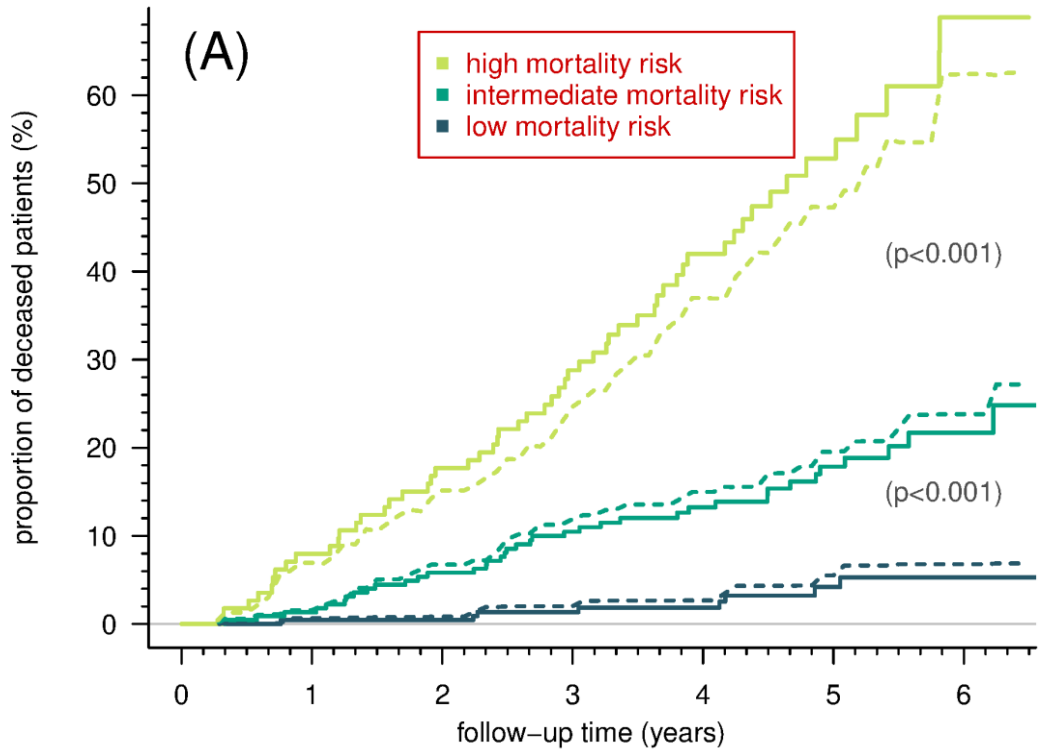

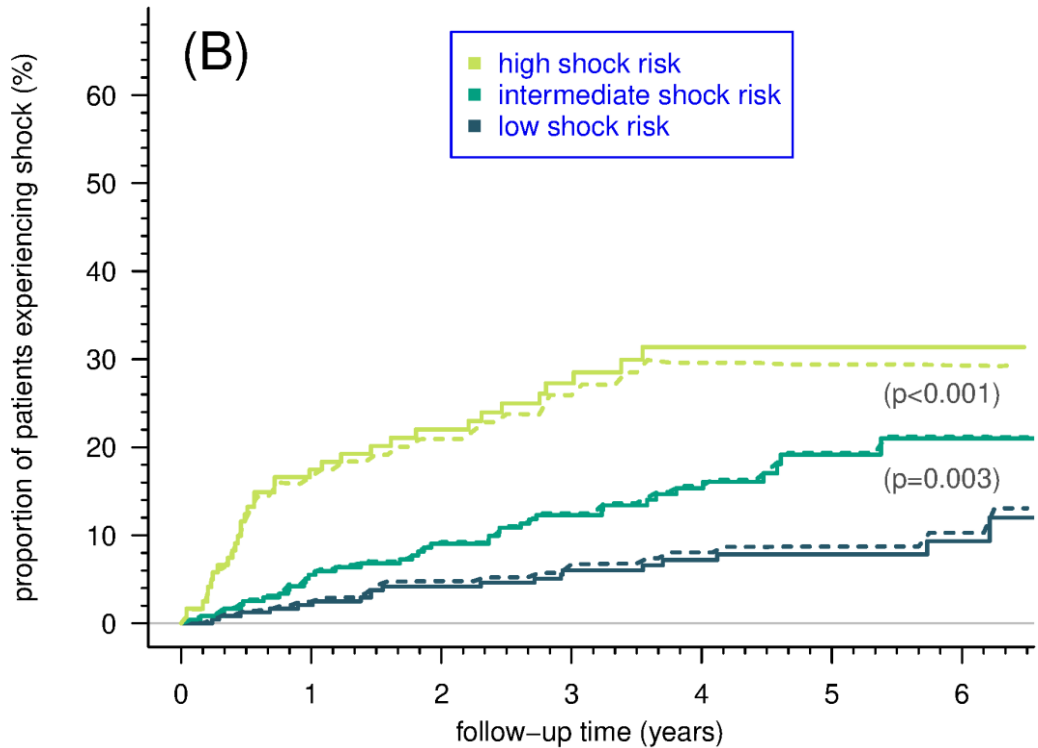

patients at risk:

| | | | | | | | |
|---|---|---|---|---|---|---|---|
| low | 241 | 233 | 226 | 192 | 149 | 90 | 48 |
| intermediate | 240 | 220 | 203 | 167 | 117 | 59 | 25 |
| high | 121 | 95 | 82 | 58 | 40 | 23 | 4 |

## Tables

**Table 1: Inclusion and Exclusion Criteria of the EU-CERT-ICD prospective study (modified from Zabel et al(20) with permission (EU-CERT-ICD design publication)**

Inclusion Criteria

- Ischemic or dilated cardiomyopathy, LVEF ≤ 35 %

- NYHA (New York Heart Association) functional class II-III (or NYHA functional class I and LVEF ≤ 30%)

- Indication for primary prevention ICD treatment according to current European Society of Cardiology guidelines

- Age ≥ 18 years

- Written informed consent

Exclusion criteria

- Secondary prophylactic ICD indication

- Planned or indicated cardiac resynchronization therapy

- Unstable cardiac condition

- Higher degree AV-block

- Previous pacemaker or device therapy

- Life expectancy ≤ 1 year

**Table 2: Parameters to be analyzed in the EU-CERT-ICD prospective study from 12-lead 24-hour Holter recordings (modified from Zabel et al(20) with permission (EU-CERT-ICD design publication)**

24 hour – Holter parameters

- Number of premature ventricular complexes

- Number of episodes and rate of non-sustained ventricular tachycardia

- Respiration triggered sinus arrhythmia

- Modified moving average T-wave alternans

- Periodic repolarization dynamics

- Heart rate variability including standard deviation of RR intervals (SDNN), root mean square of successive differences in RR intervals

- Frequency domain HRV parameters

- Heart rate turbulence: turbulence onset, turbulence slope

- Acceleration capacity

- Deceleration capacity

Advanced short-term ECG parameters

- T-peak-to-T-end interval

- J-point elevation

- Fractionation index

- Early repolarization

- Short-term variability of the QT interval

- Total cosine R-to-T

- Relative T-wave residuum

- T-wave morphology dispersion

- T-wave loop dispersion

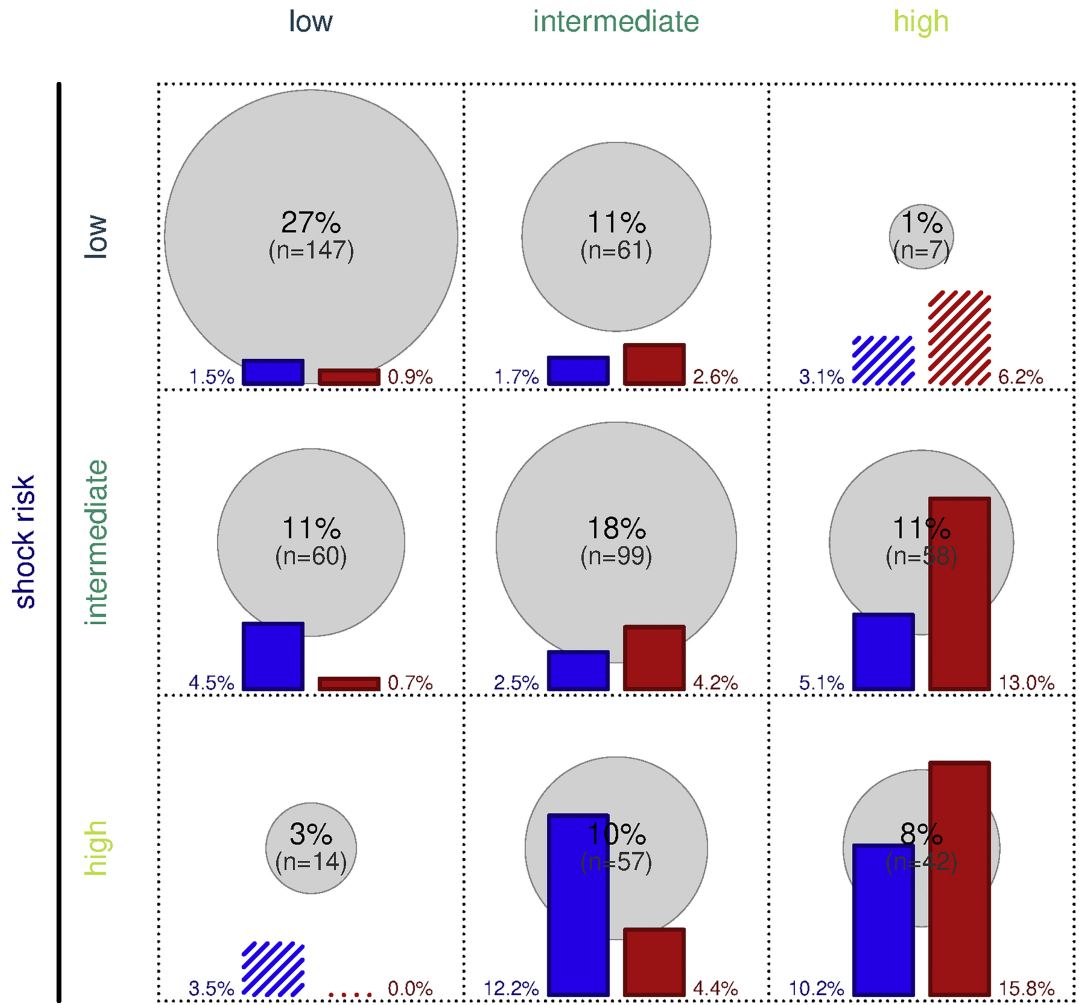

Figure 1

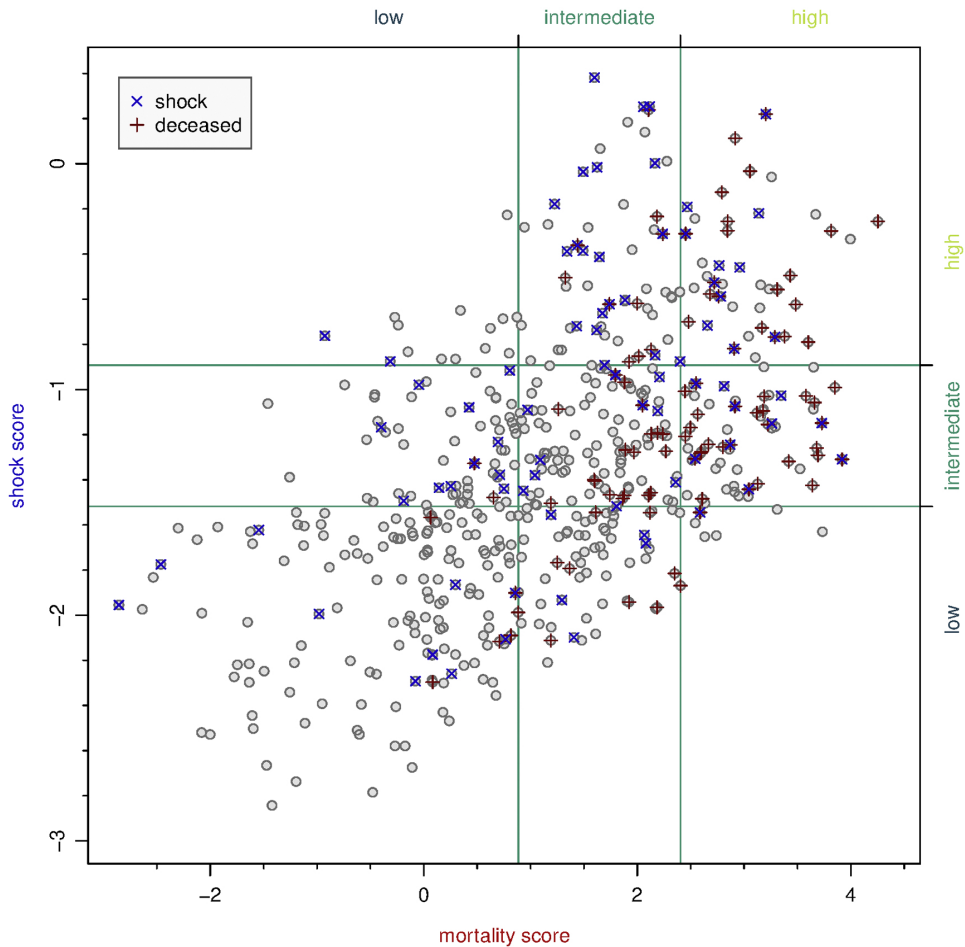

Figure 2

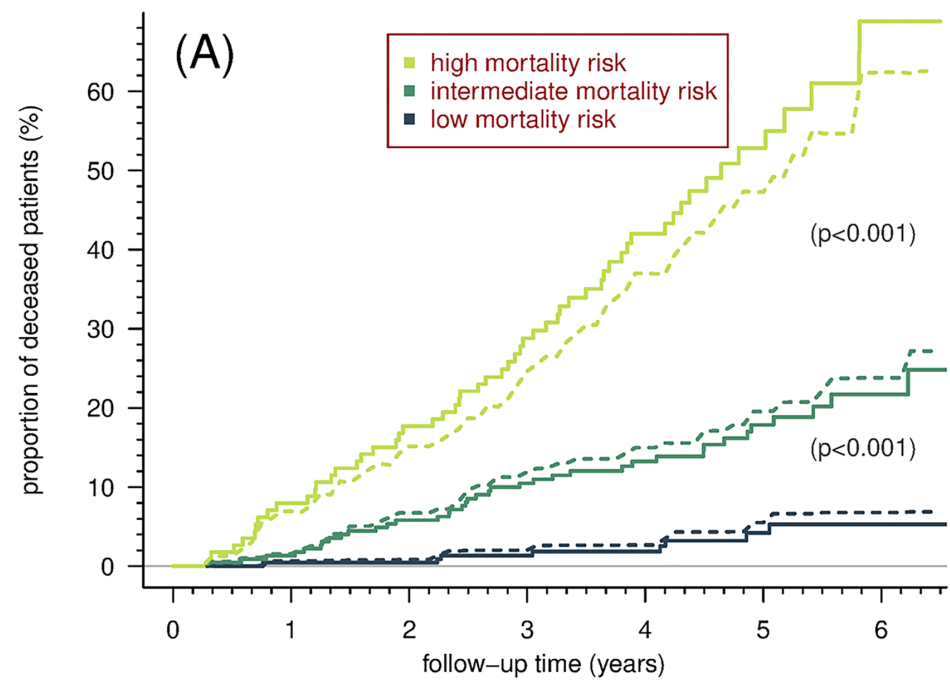

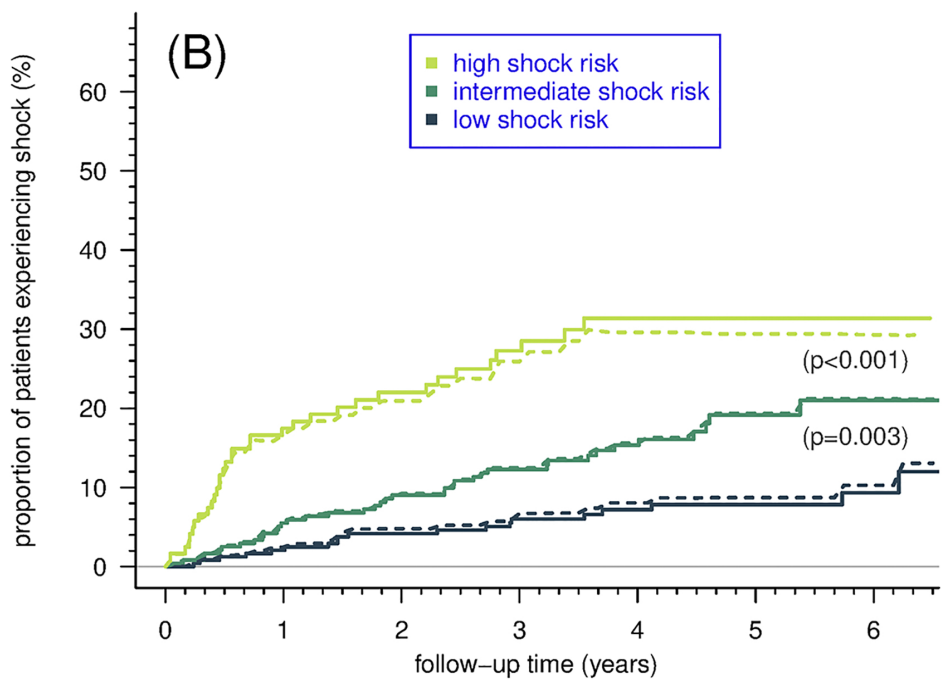

Figure 3